\providecommand{\keywords}[1]{
  \small\textbf{\textit{Keywords---}} #1
}
\providecommand{\doi}[1]{
  \small\textbf{\textit{Article doi---}} #1
}
\documentclass[a4paper,fleqn,3p]{elsarticle}
\usepackage{hyphenat}
\usepackage{graphicx}
\def\code#1{\texttt{#1}}

\begin{document}

\title{A Position and Pulse Shape Discriminant \textit{p}\hyp{}Terphenyl Detector Module}                      

\author[label1,label2]{D.P. Scriven}
\ead{scrivend@tamu.edu}
\author[label1,label2,label3]{G. Christian}
\author[label1,label2,label5]{G.V. Rogachev}
\author[label1]{C.E. Parker}
\author[label4]{L.G. Sobotka}
\author[label1]{S. Ahn}
\author[label1]{G. Chubarian}
\author[label1]{S. Ota}
\author[label1,label2,label5]{E. Aboud}
\author[label1]{J. Bishop}
\author[label1]{E. Koshchiy}
\author[label4]{A.G. Thomas}

\address[label1]{Cyclotron Institute, Texas A\&M University, College Station, Texas 77843}
\address[label2]{Department of Physics and Astronomy, Texas A\&M University, College Station, Texas 77843}
\address[label3]{Department of Astronomy and Physics, Saint Mary's University, Halifax Nova Scotia, B3H 3C3, Canada}
\address[label4]{Departments of Chemistry and Physics, Washington University, St. Louis, Missouri 63130}
\address[label5]{Nuclear Solutions Institute, TAMU 3366
College Station, TX 77843-3366 USA}

\begin{abstract}
We present the development of a neutron detector array module made with $\textit{para}$-terphenyl, a bright, fast, n/$\gamma$ discriminating crystalline organic scintillator. The module is comprised of 2 cm $\times$ 2 cm $\times$ 2 cm  $\textit{p}$-terphenyl crystals that have been optically coupled together to create a $\textit{pseudo-bar}$ module. While only relying on two photo detectors, the module is capable of distinguishing interactions between up to eight crystals. Furthermore, the module retains the $\textit{p}$-terphenyl's pulse shape discrimination (PSD) capability. Together this makes the pseudo-bar module a promising position-sensitive neutron detector. Here we present characteristics of the pseudo-bar module - its timing resolution as well as its pulse shape and position discrimination capabilities, and briefly discuss future plans for utilizing an array of pseudo-bar modules in a useful neutron detector system.
\end{abstract}

\maketitle
\noindent
\doi{https://doi.org/10.1016/j.nima.2021.165492}

\noindent
\section{Introduction}
\label{S:1_Intro}
The spectroscopy of fast neutrons is valuable for a wide range of studies in experimental nuclear physics and applied physics. Fast, position-sensitive, $n/\gamma$ discriminating , and high-efficiency neutron arrays are essential tools in modern nuclear physics. Experiments of interest include measurements of transfer reactions such as (p, n), (d, n), or (${}^{3}$He, n). Reaction measurements such as these are valuable tools to study nuclear structure away from stability and to constrain astrophysically important capture reactions. The structure of neutron-rich nuclei can also be probed through measurements of single- and multiple-neutron decay from unbound states. Neutron spectroscopy also has applications for nuclear astrophysics, e.g.\ through measurements of decay branching ratios, as well as national security applications such as those presented in Aboud \emph{et al}.~2020 (submitted to IEEE Transactions in Nuclear Science).

The primary method used for neutron spectroscopy is neutron time-of-flight ($n$TOF). Here, the time-of-flight is recorded by the neutron detector relative to a synchronized trigger detector or with the RF of an accelerator. With good timing and position resolution, one can accurately reconstruct the incident neutron kinetic energy. Most current $n$TOF detectors rely on organic (hydrocarbon) scintillators which have a high $^1$H density to facilitate $n-p$ scattering and fast rise times that allow for precise time tagging. Additionally, certain organic materials have different intrinsic decay times for the recoils of electrons (induced by $\gamma$ rays) and protons. This allows the technique of pulse shape discrimination (PSD) to be used for $n/\gamma$ particle ID. This ability is crucial in many experiments, especially those where secondary radiation fields can be expected, e.g. from rare isotope beams (RIBs), or in experiments where neutron and $\gamma$-decaying states are being populated simultaneously.

Liquid scintillators like NE-213 offer good timing, $n/\gamma$ PSD, high light output, and a high H/C ratio. However, off the shelf liquid scintillation detectors are usually manufactured in volumes which have poor inherit position resolution, and the scintillators themselves are usually toxic, volatile, and flammable. Many solid organic ("plastic") scintillators are commonly used as well. While these detectors are fast, have a good H/C ratio, and can be constructed in position-sensitive geometries, conventional plastics are not $n/\gamma$ discriminating. Recently, solid organic scintillators offering PSD capabilities have become commercially available, e.g.\ Eljen EJ-276 \cite{EJ276_DataSheet,EJ276_ZAITSEVA201288,EJ276_ZAITSEVA201897}, and $para$-Terphenyl.

One popular method of constructing position-sensitive arrays of organic scintillators is to produce detectors in a long bar made of cast plastic or liquid scintillator housed in an elongated casing. These bars typically have a photomultiplier tube (PMT) situated on either end, allowing position along the length of the bar to be deduced from the time difference between the two signals. Such bars range in size but can upwards of a meter or more in length \cite{MONA1,MONA2,LENDA_2012,NeuLand,VANDLE,ND_NWALL,SABRE}. Scintillation bars are commonly stacked in height and depth to create large, modular arrays such as MoNA \cite{MONA1,MONA2}, LENDA \cite{LENDA_2012}, NeuLAND \cite{NeuLand}, and VANDLE \cite{VANDLE}. While the time resolution ($\sim$0.2-1 ns) and PSD capabilities of such detectors are reasonable for high-quality neutron spectroscopy the position resolution is generally no better than 4-10 cm. $n$TOF energy resolution can be improved with greater position sensitivity of detector segments. One approach is taken by the NEXT project \cite{NEXT} by mating small bars of the PSD plastic EJ-276 \cite{EJ276_DataSheet,EJ276_ZAITSEVA201288,EJ276_ZAITSEVA201897} to a multi-anode (and thus position-sensitive) PMT. This state-of-the-art project has achieved excellent results in the spatial-PID-timing space as does the approach described in the present work. 

In order to improve upon present neutron detector arrays further, we developed a method for producing pseudo-bar-style detector modules made from the solid organic scintillator $para$-terphenyl. $P$-terphenyl exhibits exquisite PSD, comparable to many liquid scintillators \cite{Yanagida_2014}, and an appreciable light output of up to 3.5 $\times10^4$ photons/MeVee \cite{Angelone_2014} - a factor $\sim$2-3 brighter than commercially-available.

$P$-terphenyl is grown in small cylinders with a maximum diameter of ${\sim} 5$ cm and maximum height of ${\sim} 30$ cm. Because of size limitations from the crystal growth process, it isn't possible to manufacture bars with lengths comparable to typical plastic or liquid scintillator bars. To circumvent these limitations, we have developed a novel pseudo-bar design, wherein small cubic elements of $p$-terphenyl, cut from cylindrical crystal growths, are coupled together using standard optical grease to form an elongated bar-like geometry. We place a PMT on either end of the pseudo-bar to read out the resulting scintillation light. As explained in section 3, this scheme allows the cube of interaction to be uniquely identified down to low thresholds.

In the present work, we show the results of testing 6-crystal pseudo-bar modules, with each crystal being 2 cm $\times$ 2 cm $\times$ 2 cm in size. We begin the discussion with the results of exploratory simulations $\S$\ref{S:2_Sim}, followed by the construction of a detector module $\S$\ref{S:3_Construction}, position discrimination $\S$\ref{S:4_Position}, $n/\gamma$ discrimination $\S$\ref{S:5_PSD}, the timing resolution $\S$\ref{S:6_Timing}, and conclusion $\S$\ref{S:7_Conclusion}.

\section{Simulation}
\label{S:2_Sim}
The final goal of this project was to create a large, pixelated neutron detector array with a high degree of segmentation for good neutron energy resolution.  Before beginning detector construction, exploratory simulations were done to investigate the relationship between kinetic energy resolution and voxel size, for a range of incident neutron energies. We describe the 3-dimensional scintillator pixels as volumetric pixels, $voxels$. The results were used to choose a voxel size that optimally balances energy resolution with cost and complexity. Shown in Table \ref{tbl1} are the range of neutron energies and voxel sizes used in these simulations. All possible combinations of these parameters were explored.

\begin{table}
\caption{A list of various parameters used in Geant4 simulations.}
\label{tbl1}
\begin{tabular*}{0.5\linewidth}{r|rrrrrrr}

$T_{n, inc}$ [MeV] & 1 & 5 & 10 & 15 & 20 & 25 & 30 \\
Voxel Size [mm] && 5 & 10 & 20 & 30 & 50 & 100 \\
 
\end{tabular*}
\end{table}

Using Geant4 \cite{Geant4}, a detector wall was generated with dimensions of width and height, $w=h=1$ m, and thickness $t=0.1$ m. The volume is filled with voxels of the dimensions shown in Table \ref{tbl1}. For simplicity, no spaces are left for photo readout, electronics, or support structure. Mono-energetic neutrons were fired from a point-source with a flight path of 1 m.  The Geant4 default $n$-$p$ and $n$-$^{12}$C cross sections have been replaced using the \code{MENATE\_R} code for cross section generation \cite{menate1,menate2,menate3}. The \code{MENATE\_R} code more accurately reproduces neutron scattering and reaction cross sections for neutrons with $T_n$ of $1-300$ MeV. These cross sections give the proper treatment for $^{12}$C inelastic scattering and neutron-induced breakup which are important contributors to sub-threshold events that alter the incident neutron's trajectory ("dark scattering").

We take the detection position to be the center of each voxel, and apply a 200 keV threshold. The detected position and time-of-flight of the earliest hit are used to calculate the measured neutron incident kinetic energy, $T_n$. Histograms of the difference between $T_n$ and the incident neutron energy are analyzed to determine the resulting energy resolution. These simulations use a timing resolution of $\sigma_{timing} = 255$ ps (FWHM=600 ps). This is based on realistic time resolution discussed in $\S$\ref{S:6_Timing}.

\begin{figure}
    \centering
	\hspace{0cm}\includegraphics[width=0.6\columnwidth]{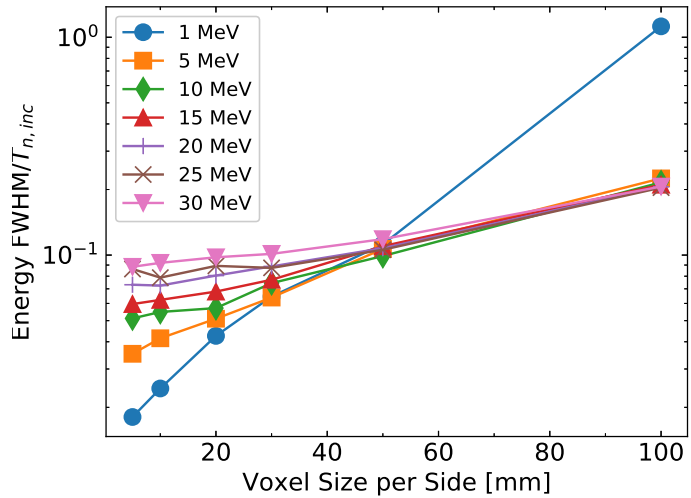}
	\caption{The resolution of reconstructed neutron kinetic energy as a function of the incident energy, shown for the full range of crystal sizes tested. The timing resolution is taken to be 600 ps FWHM (255 ps $\sigma$).}
	\label{fig:Eres}
\end{figure}

In Fig. \ref{fig:Eres}, we show the energy resolution for various crystal sizes and incident energies. We focused on the variation of energy resolution with voxel size, in order to guide our choice of the optimal crystal geometry. As expected, the resolution improves with decreasing voxel size. Except for the lowest energy neutrons, we observe a plateau in the resolution gains for voxel sizes below 20 mm. This analysis guided the choice of voxel size for our prototype module. After considering the cost per unit volume (optimal at about 30 mm), delivery time (increases with size), and the resolution issues, we set upon 20 mm crystals.

\section{Construction and Setup}
\label{S:3_Construction}
With the goal of building a large, thick neutron detector of dimensions similar to those simulated, light transport techniques would be required in order to keep the scintillator densely packed. Some effort was spent using wavelength shifting scintillator slabs in an optically multiplexed configuration to pipe light from the interior of the $p$-terphenyl matrix to PMTs around the perimeter; however, these efforts were unsuccessful. This led to the testing and development of pseudo-bar modules. 

\begin{figure}
    \centering
	\hspace{0cm}\includegraphics[width=0.6\columnwidth]{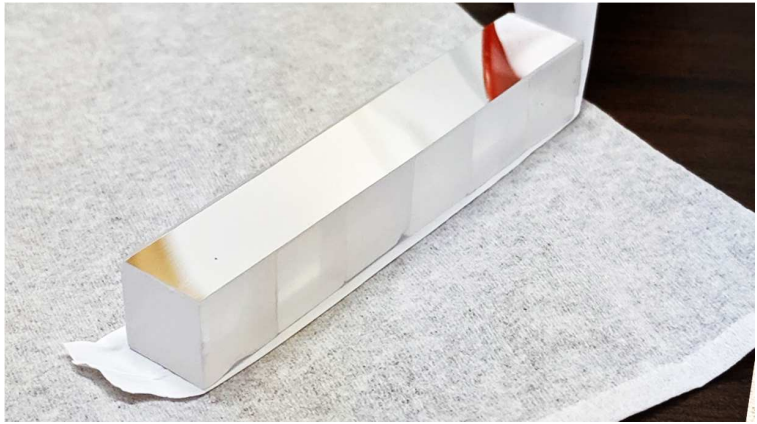}
	\caption{Photograph of a pseudo-bar in the midst of assembly. The six bare $p$-terphenyl crystals are shown with the specular ESRF on the top side of the bar.}
	\label{fig:pseudo-bar}
\end{figure}

A pseudo-bar is constructed by optically coupling six $p$-terphenyl crystals together with Eljen EJ-550 Optical Grade Silicone Grease. A pseudo-bar with 8 crystals were also tested. While reasonable performance was also obtained, the slightly shorter six-element detector, had superior performance. The six-crystal scintillator bar is covered lengthwise  on all four sides by a strip of 3M$\textsuperscript{TM}$ Vikuiti$\textsuperscript{TM}$ Enhanced Specular Reflective Film (ESRF). The scintillator and ESRF is further wrapped in white Teflon while leaving the two ends of the scintillator bar exposed. The Teflon is used primarily to hold the scintillator bundle together, and to ensure that the ESRF is pressed tightly against the crystals. The wrapped scintillator is placed into a 3-sided aluminum housing which has an access port at each end with an O-ring lining the radius of the port. Inside the port, a hollow $\mu$-metal draw-tube is inserted and held in place by the O-ring. The $\mu$-metal is to neutralize the effects of magnetic fields on the PMTs, should pseudo-bars be deployed in the stray magnetic field of a spectrometer for instance. The two ends of the scintillator bar are each optically coupled with EJ-550 to a Hamamatsu R1450 PMT which are inserted into the draw-tube of the housing. The top of the housing is capped off, and small amounts of black electrical tape are placed around contact points to ensure light tightness. A photograph of the un-wrapped pseudo-bar in the midst of assembly is shown in Fig. \ref{fig:pseudo-bar}.

The emission spectrum of $p$-terphenyl peaks at 420 nm, which is nearly matched to the wavelength of peak quantum efficiency (QE) of the PMTs. The EJ-550 has a refractive index of 1.46 which has a small mismatch with the $p$-terphenyl, having a refractive index of 1.65. The 3M$\textsuperscript{TM}$ ESRF has a reflectance of $>98\%$ in the visible range.

Testing of the pseudo-bar made use of both digital-signal-processing (DSP) technology and standard analog signal processing techniques. Using both allowed us to exploit the advantage of each to more fully understand the device. We will first describe the tests that employed the DSP.

A CAEN V1730 waveform digitizer with 500 MHz temporal sampling rate and a 14-bit dynamic range internal ADC was used to record list mode data and also waveforms for further offline analysis. The digitizer uses DSP PSD  firmware which calculates two user defined integrals over each waveform. These integrals are calculated online and are recorded in list mode denoted as $E_{long}$ and $E_{short}$. $E_{long}$ is used for the total energy (from a single PMT) while, the two integrals together are used for PSD, as explained in \S\ref{S:5_PSD}. The digitizer firmware also generates an online CFD for timing, and the timestamp is also recorded. We use an in-house code to match events as coincidences between the two PMTs. 

The PMTs are gain matched using a $^{137}$Cs source, while being operated with a rather large voltage of between 1500-1700 V. A $^{207}$Bi source is used in addition to the $^{137}$Cs for energy calibration. The $p$-terphenyl, being a low-Z hydrocarbon, is insensitive to the photo-peak from these sources, so instead the Compton edge is used for energy calibration. The $n/\gamma$ PSD properties are characterized using a standard $^{252}$Cf $n/\gamma$ source. 

\begin{figure}
    \centering
	\hspace{0cm}\includegraphics[width=0.6\columnwidth]{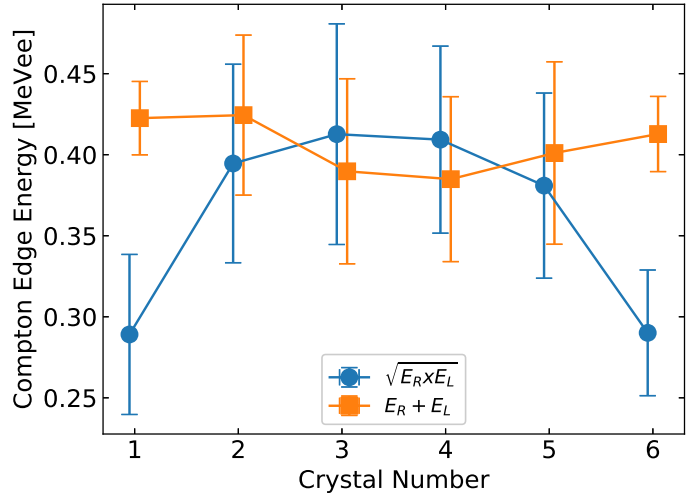}
	\caption{Comparison of the arithmetic sum ($E_L+E_R$) and geometric sum ($\sqrt{E_L\times E_R}$) methods of calculating the total energy. The plot shows the peak positions and $1\sigma$ widths of the 394 keV Compton edge originating from a $^{207}$Bi source, as a function of crystal number. The data points show the central values of each peak, while the error bars show each peak's $1\sigma$ width.}
	\label{fig:Compton}
\end{figure}

\begin{figure}
    \centering
	\hspace{0cm}\includegraphics[width=0.6\columnwidth]{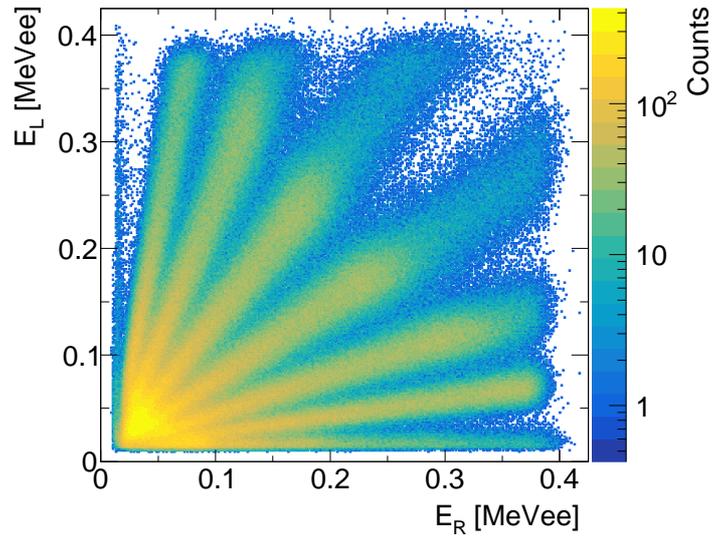}
	\caption{The integrated total energy from both left and right PMTs plotted against each other event-by-event using a $^{207}$Bi source. The two less prominent bands which straddle the abscissa and ordinate come from noise.}
	\label{fig:EE}
\end{figure}

In this detector, the energy calibration is performed by taking the arithmetic sum of the energy measured by each PMT (left and right), such that, $E_{total} = E_{L}+E_{R}$. Conventional bar-type detectors use $E_{total}\sim \sqrt{E_L \times E_R}$ approach to calculate total energy due to the exponential attenuation along the length of the bar. Experimental evidence suggests that for this pseudo-bar however, the attenuation is small compared to the reflection of light at the interfaces between $p$-terphenyl crystals, and the dominant effect is a reflection of light on the surfaces separating the $p$-terphenyl crystals due to difference between the index of refraction of the crystals and the optical grease (1.64 and 1.46 respectively). In this case, the $E_{total}$ calculated as an arithmetic sum of the left and right PMTs is approximately independent of the firing crystal's location along the bar, whereas the $\sqrt{E_L \times E_R}$ is a function of the crystal's location. This can be seen in Fig. \ref{fig:Compton} that shows the Compton edge of a 570 keV $\gamma$ ray as a function of each crystal's location along the bar calculated using the two different methods for determining the $E_{total}$. In Fig. \ref{fig:EE} we show the equipartition of scintillation light between the two PMTs that exemplifies the boundary driven attenuation in the detector. Due to low Z of the scintillator material, the photoelectric effect cross section is small and Compton scattering dominates. In the figure, the primary $\gamma$ recoils seen are of energy 393 keV and 858 keV, originated by 570 keV and 1064 keV $\gamma$ rays respectively. 

\section{Position Discrimination}
\label{S:4_Position}
As previously stated, one key goal of this project was to develop an optical readout method which lets one identify the crystal of interaction reliably. In the detector, we observe light loss at each crystal interface due to reflection. This becomes a beneficial feature for the pseudo-bar module, leading to position sensitivity between the six crystals despite only end photon readouts.

\begin{figure}
    \centering
	\hspace{0cm}\includegraphics[width=0.6\columnwidth]{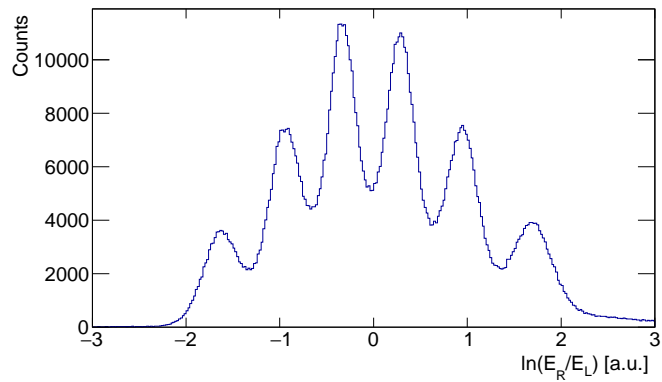}
	\caption{The position spectrum as calculated from the energy of the left and right PMTs. Here, each peak in the distribution corresponds to a single crystal.}
	\label{fig:RL}
\end{figure}

By looking at coincident signals from the "left" and "right" PMTs, one easily notices features in spectra that are indicative of position discrimination. This is demonstrated in Fig. \ref{fig:EE}, where each of the six bands observed in the plot of $E_L$ vs. $E_R$ corresponds to light originating from a different crystal. This spectrum can be condensed into a single parameter, $\ln(E_R/E_L)$, which is used to discriminate between crystals. The spectrum of $\ln(E_R/E_L)$ is shown in Fig. \ref{fig:RL}. A one dimensional Monte Carlo calculation that simulated reflection at crystal boundary crossing was performed. The six bands in the Fig. 5 distribution were each fit linearly. The corresponding curves were compared to results of the Monte Carlo and most closely matched a light reflectance factor of 0.4 at each boundary.

\begin{figure}
    \centering
	\hspace{0cm}\includegraphics[width=0.6\columnwidth]{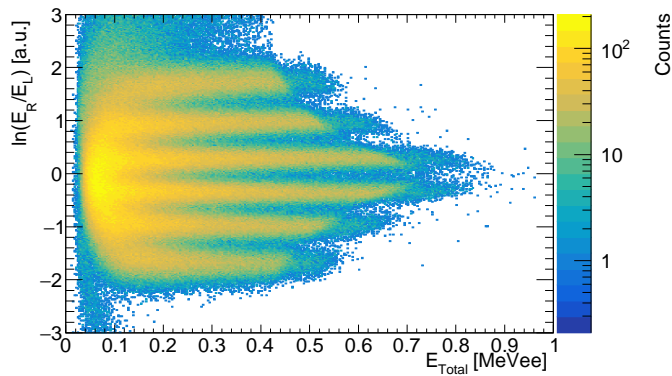}
	\caption{The 2D histogram showing $\ln(E_R/E_L)$ vs $E_{total}$ from a $^{252}$Cf source. Both neutrons and $\gamma$ ray events are shown. It is easily seen that there is an energy dependence in the discrimination between each crystal.}
	\label{fig:RL_E}
\end{figure}

The six peaks in the spectrum of Fig. \ref{fig:RL} are not perfectly separated at all energies with
discrimination being worse for lower energies. This is evident in Fig. \ref{fig:RL_E} which shows the energy dependence of the position separation. We use the left-to-right convention for crystal designation; starting from the left they are labelled C1-C6. To quantify the crystal distinguishability a figure of merit (FOM) is defined as,
\begin{equation}
\label{FOMx}
FOM_{i,\ i+1} = \frac{\mu_i-\mu_{i+1}}{2.355(\sigma_i+\sigma_{i+1})}.
\end{equation} 
Here, the indices are for the $i^{\textnormal{th}}$ peak, and its neighbor, the $i^{\textnormal{th}}+1$, where $i$ ranges from 1 to 5, with $\mu$ and $\sigma$ being the mean and standard deviations of fitted Gaussian distributions. This FOM describes how well separated any two Gaussian fits are, based on the width of the Gaussian and the separation of the means along the abscissa. With this metric,  $FOM=1$ corresponds to an overlap of the two Gaussian functions at the $2.355\sigma$ level. A $FOM>1$ describes an overlap of the Gaussian functions further than $2.355\sigma$ away, while $FOM<1$ is for a crossing at less than $2.355\sigma$. 

\begin{figure}
    \centering
	\hspace{0cm}\includegraphics[width=0.6\columnwidth]{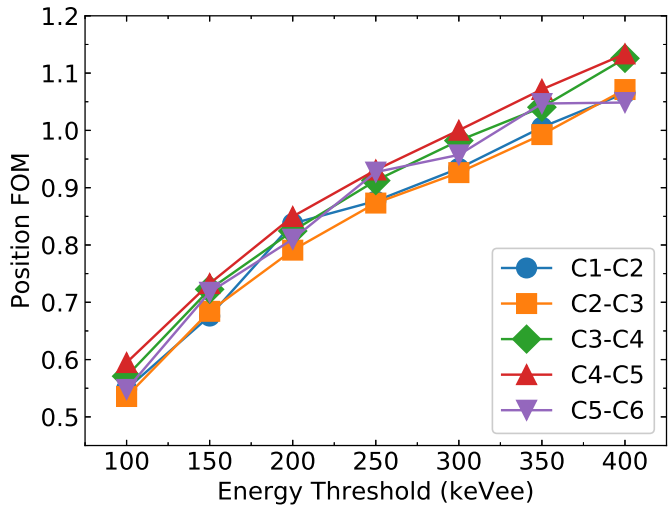}
	\caption{The position FOM profiles for each pair of neighboring crystals in a pseudo-bar.}
	\label{fig:FOMx}
\end{figure}

From the spectrum in Fig. \ref{fig:RL_E}, we apply an energy window 50 keVee wide and calculate the FOM for each neighboring pair of Gaussian fits.  By stepping the window and re-evaluating the FOM we generate FOM profiles which graphically describe how the position discrimination changes as a function of energy threshold. The FOM profiles for position discrimination are shown in Fig. \ref{fig:FOMx} for each neighboring pair. As is shown by FOM profiles, the position separation clearly depends on the energy threshold, and on the energy. A $FOM=1$ is reached near 300 keVee. Only thresholds $\geq100$ keVee were considered; below this, the neighboring peaks become un-resolved and a reliable FOM could not be determined.

\section{Pulse Shape Discrimination}
\label{S:5_PSD}
$P$-terphenyl has been shown to exhibit PSD that is competitive with other solid-state, neutron-sensitive materials as shown in Refs. \cite{Yanagida_2014, Angelone_2014}. By mating one crystal of $p$-terphenyl to a PMT, ordinarily one observes exquisite $n/\gamma$ PSD. In this more complex detector module, light transport is haltered, giving rise to the position discrimination properties shown in $\S$\ref{S:4_Position}. Here we verify that the $n/\gamma$ PSD is still viable, despite the less-conventional light transport.

A key property for any neutron detector is the ability to distinguish between neutrons and $\gamma$ rays as the latter are nearly always present with the former. Many scintillating materials (like $p$-terphenyl) are capable of this particle discrimination. This is possible because the proton (electron) recoils resulting from neutron ($\gamma$ ray) interactions generate pulses that differ in their decay times. Thus by integrating the pulses over different time windows, one is able to determine the type of incident radiation. Pulse shapes are discriminated by assigning a PSD value resulting from these integrals.

Our PSD is done with the standard tail-over-total method utilizing the $E_{long}$ and $E_{short}$ values integrated by the DSP (or offline). In this method, the PSD is calculated by 
\begin{equation}\label{eq:PSD}
PSD = \frac{E_{long}-E_{short}}{E_{long}},\end{equation}where the numerator gives the expected value of the tail integration. A pre-gate which defines the start of the integration before the CFD trigger is established. With the pre-gate of 60 ns, we found that the long and short gates which optimized the PSD were 150 ns and 30 ns, respectively.

\begin{figure}
    \centering
	\hspace{0cm}\includegraphics[width=1.\columnwidth]{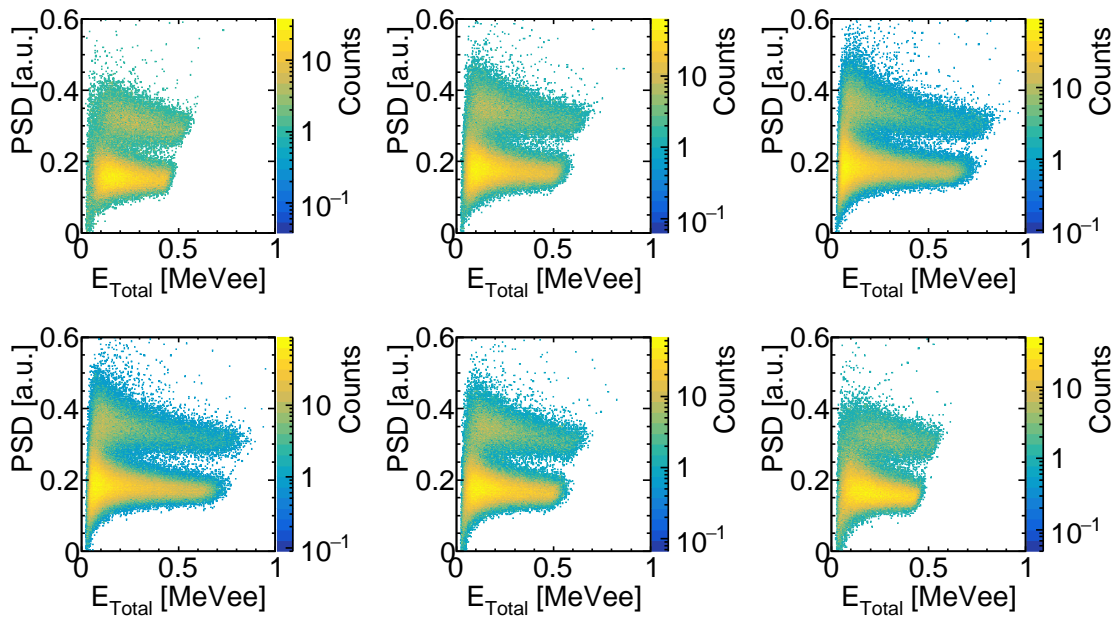}
	\caption{PSD vs $E_{total}$ for each crystal in the detector. From top left to bottom right are crystals C1-C6 in that order. The PSD shown is saturated because of a known problem in the CAEN V1730 digitizer short gate which truncates events with sufficiently large short integrals.}
	\label{fig:PSD_Panel}
\end{figure}

As was discussed in $\S$\ref{S:4_Position}, we are able to gate on the individual crystals within the pseudo-bar module. By making these gates, we are able to look at other parameters such as the PSD, position wise. In Fig. \ref{fig:PSD_Panel} we show the PSD in each crystal of the detector as a function of $E_{total}$. 

\begin{figure}
    \centering
	\hspace{0cm}\includegraphics[width=0.6\columnwidth]{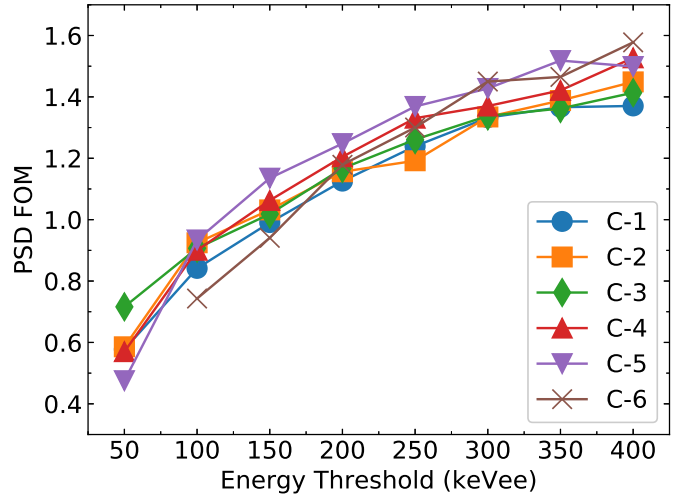}
	\caption{The PSD FOM profiles for each pair of neighboring crystals in a pseudo-bar.}
	\label{fig:FOMpsd}
\end{figure}

We define now for the $n/\gamma$ discrimination a figure of merit which takes the same form as that used for position discrimination in Eq. \ref{FOMx}.
\begin{equation}
\label{FOMng}
FOM_{n/\gamma} = \frac{\mu_n-\mu_\gamma}{2.355(\sigma_n+\sigma_\gamma)}
\end{equation}
The subscripts $n$ and $\gamma$ correspond to mean and $\sigma$ of the neutron and $\gamma$ ray peaks in the 1-D PSD spectrum. We again generate FOM profiles (as in $\S$\ref{S:4_Position}). The profiles are shown in Fig. \ref{fig:FOMpsd}. Below a 50 keVee threshold (100 keVee for crystal 6), the peaks were unresolved and a FOM could not be determined. One side-effect of the pseudo-bar construction is that while the outer crystals have better PSD in terms of setting a low threshold, these crystals also have an intrinsic threshold below which the end-end coincidence efficiency drops due to poor light collection on the far side. One side effect of running the PMTs at the high voltage we use, is that waveforms in the waveform digitizer can become saturated. The saturation can be seen at the high energy end of the spectra in Fig. \ref{fig:PSD_Panel}. Crystals at the end of the bar suffer the most from saturation because of their direct proximity to one of the PMTs. These crystals suffer less light attenuation while scintillation light throughput from central crystals is decreased.

\begin{figure}
    \centering
	\hspace{0cm}\includegraphics[width=0.6\columnwidth]{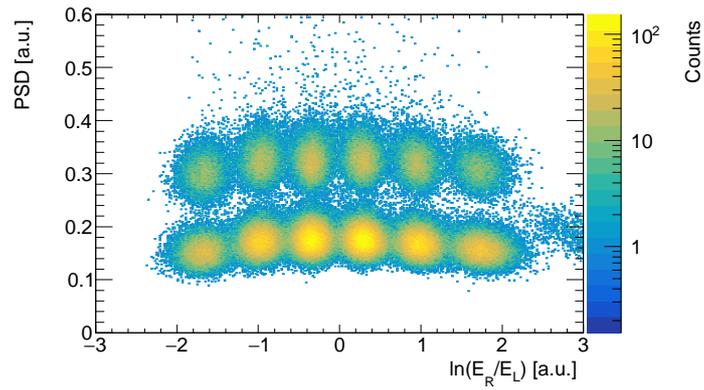}
	\caption{PSD vs. position parameter $ln(E_R/E_L)$. Here, the top band of distributions corresponds to neutrons, while the bottom band of distributions are $\gamma$ rays. Each column of two distributions corresponds to a crystal. Showing only events with $E_{total}$ above 300 keVee it becomes easy to separate neutrons in from $\gamma$ rays and also neutron events between each crystal.}
	\label{fig:PSDRL}
\end{figure}

In Fig. \ref{fig:PSDRL} the $n/\gamma$ discrimination FOM is plotted versus the position dispersive variable $ln(E_R/E_L)$. For this figure an energy threshold of 300 keVee is applied. High-quality PID is achieved for each crystal. From Fig. \ref{fig:FOMpsd} we are able to declare a $n/\gamma$ PSD threshold of 150 keVee.

\section{Timing Resolution}
\label{S:6_Timing}
The $n$TOF method allows a particle's kinetic energy to be measured by using the standard kinetic energy formula, $T=m(d/t)^2/2$ (assuming non-relativistic neutrons). Therefore, the energy resolution is heavily dependent on the position resolution and timing resolution. In our detector, the position resolution is essentially fixed to the crystal size, and here we discuss the results of measurements to determine the timing resolution. 

To understand the timing resolution, we setup a pair of ultra-fast cesium fluoride (CsF) scintillation $\gamma$-ray detectors (25 mm in diameter by 40 mm deep) in conjunction with the pseudo-bar module.  The two CsF detectors are described as CsF$_{start}$ and CsF$_{stop}$. With a $^{22}$Na source on the bench, the CsF$_{start}$ is placed opposite of CsF$_{stop}$, each equidistant from the source at 20 cm. A pseudo-bar module is placed at the location of CsF$_{stop}$, also acting as a stop. The $^{22}$Na decay produces positrons which annihilate with local electrons to create coincident 511 keV $\gamma$ rays which are emitted back to back (180$^{\circ}$ apart) from the annihilation site. The two $\gamma$ rays are then observed coincidentally in the CsF$_{start}$-CsF$_{stop}$ or the CsF$_{start}$-pseudo-bar system. 

\begin{figure}
    \centering
	\hspace{0cm}\includegraphics[width=0.9\columnwidth]{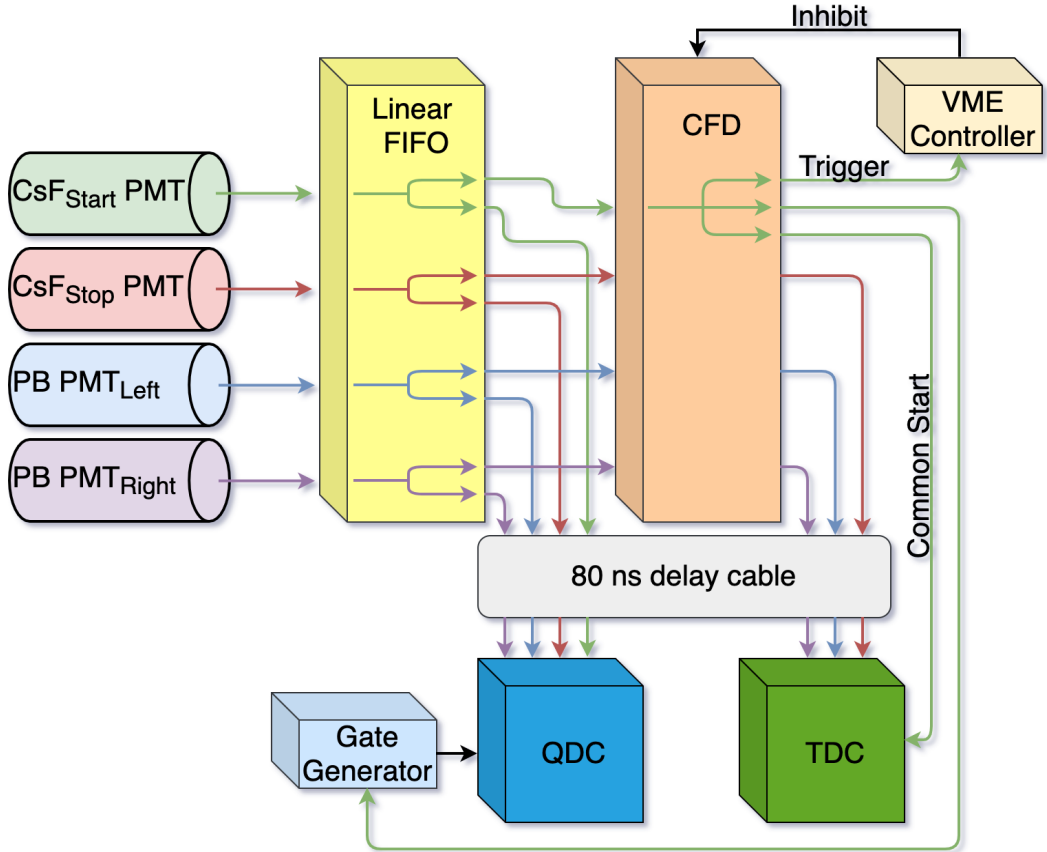}
	\caption{A diagram showing the DAQ schematic used during the timing measurement using conventional electronics.}
	\label{fig:electronics}
\end{figure}

In $\S$\ref{S:4_Position} and $\S$\ref{S:5_PSD}, we relied on digital data acquisition methods (DAQ). For position and PSD, the digital acquisition worked very well. The CsF however, having an intrinsic rise time of 0.35 ns  \cite{CsF_Moszynski1981271}, is too fast for the CAEN V1730 unit which has a 500 MHz sampling rate (2 ns time buckets). This sampling rate is insufficient to sample the rise of the CsF waveform, making time tagging unreliable in both leading edge or constant fraction discrimination modes. This lead us to approach the timing resolution measurement with analog electronics. A full electronics schematic is shown in Fig. \ref{fig:electronics}.

Raw signals from the CsF and pseudo-bar PMTs are split using a Phillips Scientific 740 quad linear fan-in/fan-out. These signals are sent, one each, to CAEN V792 VME QDC for integration and energy measurement, and also to an Ortec CF8000 constant fraction discriminator. The CFD module provides again, two outputs, one which is sent to a CAEN V775 TDC for time logging, and the CsF logic is sent to a gate generator. The Phillips Scientific 794 gate generator generates a gate for each independent PMT signal. We use a LeCroy 429 Logic Fan-in/Fan-out unit to split logic signals. In measurement mode, only the CsF CFD logic is used for triggering the DAQ. The split CsF and pseudo-bar signals and CFD logic are delayed before entering the QDC and TDC respectively, in order to fit within the integration gate. The QDC and TDC are readout via a VME controller in our CAEN VME crate.

The pseudo-bar module and CsF$_{stop}$ detector sit behind two lead bricks which are used to collimate the $\gamma$ rays, primarily for the purpose of illuminating one crystal at a time in the detector module. The bricks are placed slightly less than 2 cm apart and are $\sim2$" in thickness. This allows for single crystal analysis without the energy-dependent uncertainties that might arise from the $E_L/E_R$ position discrimination technique. 

\begin{figure}
    \centering
	\hspace{0cm}\includegraphics[width=0.6\columnwidth]{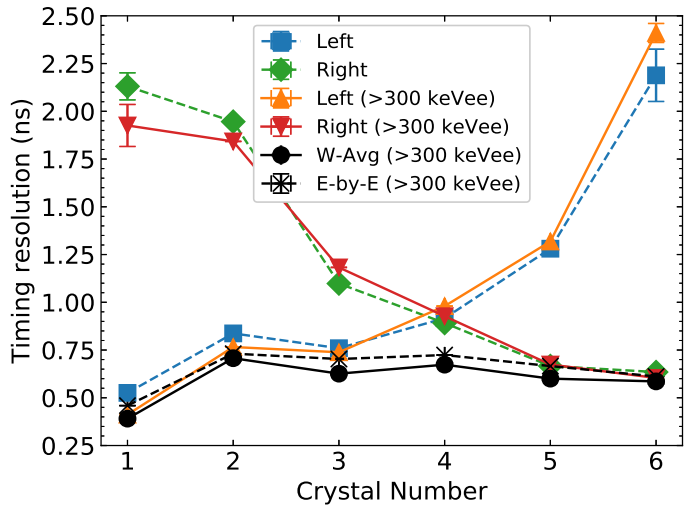}
	\caption{The timing resolution measured for each crystal. It is shown from the perspective of both left and right PMTs. "Left" and "Right" timing value are measured using all data which have energy in the timing-linearity region (where possible). Those labeled with a 300 keVee cut are measurements with only events above this energy threshold. Error-bars are produced from the errors of the fits and are multiplied by 10x for visibility. For points where no error bar is shown, errors are smaller than the data points.}
	\label{fig:timing}
\end{figure}

During analysis, we first look at the timing resolution of the CsF detectors. By, gating on 511 keV photo-peak in both CsF$_{start}$ and CsF$_{stop}$, the timing spectrum can be generated without any dependence on the signal amplitude. A time calibrator is used to determine the calibration, and a sampling resolution of $35^{+0.6}_{-0.2}$ ps/channel is found which is consistent with the nominal TDC slope. The two CsF detectors are assumed to be identical devices, and so their individual timing resolutions are determined by the deconvolution, $\sigma_{CsF} = \sigma_{start-stop}/\sqrt{2}$. Converting to a full width half maximum by $\tau_{CsF} = 2\sqrt{2ln{2}}\sigma_{CsF}$, we find a timing resolution of $\tau_{CsF} = 364$ ps.

\begin{table}
\caption{Tabulated measurement values for the timing resolution of a single pseudo-bar module. These results are shown graphically in Fig. \ref{fig:timing}.}
\label{tbl2}
\begin{tabular*}{\linewidth}{@{} lllll @{}} 
Crystal ID & $\tau_{Left}$ [ps] & $\tau_{Right}$ [ps] & $\tau_{W-Avg}$ [ps] & $\tau_{E-by-E}$ [ps] \\
C1 & 411 & 1930 & 391 & 459\\
C2 & 766 & 1840 & 707 & 731\\
C3 & 738 & 1180 & 626 & 703\\
C4 & 980 &  927 & 673 & 724\\
C5 & 1320 & 674 & 600 & 666\\
C6 & 2410 & 604 & 586 & 611\\
\end{tabular*}
\end{table}

Pseudo-bar events above the nominal position discrimination threshold of 300 keVee were used in the timing analysis as well as a repeated analysis with no energy cut. With these cuts, the widths of the timing distributions are fit by Gaussian distributions, and converted into ns using the slope determined with the time calibrator data. The distribution is then deconvoluted by $\sigma_{pseudo-bar} = \sqrt{\sigma_{dis}^2 - \sigma_{Csf}^2}$ to remove the contribution of the CsF start trigger detector. Again, converting to a FWHM, we find the timing resolution for each crystal within the detector. These measurements are tabulated in Table \ref{tbl2}, and are also shown graphically in Fig. \ref{fig:timing}. The arithmetic mean of FWHM of all six crystals is 597 ps.

Additionally, we calculate a weighted average of the
left and right time values, where the weights are the $1\sigma$ widths of the coincidence peaks. This new weighted average is calculated as \begin{equation}\sigma_{Avg} = ({\Sigma_i w_i})^{-1/2} =\sqrt{\frac{\sigma_L^2\sigma_R^2}{\sigma_L^2 + \sigma_R^2}}.\end{equation} This weighted average is also shown in Fig. \ref{fig:timing} as well as Table \ref{tbl2}. This weighted average represents the achievable timing resolution of the device using left-right coincidence timing.

In addition to the weighted average timing method, we also calculate the timing FWHM with an event-by-event analysis. The prescription here is to remove the time information between each PMT relative to the CsF$_{start}$ trigger and then calculated a weighted time average between the two pseudo-bar PMTs. This event-by-event left-right PMT weighted average time is found by, \begin{equation}t_{E-by-E} = \frac{t_L\sigma^{-2}_{L}+t_R\sigma^{-2}_{R}}{\sigma^{-2}_{L}+\sigma^{-2}_{R}},\end{equation} where the $\sigma$ values are extracted from the histograms accumulated across the entire run. This analysis more accurately reflects the method that time will be measured during an experiment when doing $n$TOF for invariant- or missing-mass analysis. The FWHM of this distribution is extracted for each crystal in the module. The results from this method are also tabulated and shown graphically Table \ref{tbl2} and Fig. \ref{fig:timing} respectively. The arithmetic mean from this measurement is 649 ps.

\section{Conclusion}
\label{S:7_Conclusion}
In summary, we have developed a novel method for constructing a bar-style neutron detector using segments made of the organic scintillator $p$-terphenyl. Containing six crystals and a PMT at each end, a single pseudo-bar module is capable of discerning crystal position confidently for events depositing as little as 300 keVee in energy. Along with the position discrimination, $n/\gamma$ pulse shape discrimination is also possible with FOM>1 in each crystal down to 150 keVee. The timing resolution of the detector is not worse than 0.707 ns for any crystal, and as good as 0.391 ns. 

More detailed simulations of these detector modules and detector arrays are planned using Geant4 incorporating the \code{MENATE\_R} package and optical photon transport. The 6-crystal modules will be built into a versatile geometry array with the intent to measure neutrons between 2-10 MeV which are common neutron energies from reactions in the MARS facility at the Texas A\&M Cyclotron Institute. For the planned commissioning experiment, the proposed array would consist of eight bars per layer, and five layers deep along the beam axis. Accounting for quenching \cite{Sardet2015} and the position threshold extracted in this article (300 keVee), we also estimate an efficiency between 20-35\% for neutrons between 2-10 MeV. Both the prototype and full array will use PSD capable integrated circuits \cite{PSD_IC} which will be capable of handling the large channel load. The prototype will be commissioned at the Texas A\&M University Cyclotron Institute with rare isotope beams produced by the Momentum Achromat Recoil Spectrometer MARS \cite{MARS}.

\section{Acknowledgements}
\label{S:8}
The authors acknowledge that this material is based upon their work supported by the U.S. Department of Energy, National Nuclear Security Administration through the Center for Excellence in Nuclear Training and University Based Research (CENTAUR) under grant No. DE-NA0003841. Additional support at Texas A\&M University is provided by the U.S. Department of Energy, Office of Science, Office of Nuclear Science, under Award No. DE-FG02-93ER40773, and at Washington University in St. Louis by by the U.S. Department of Energy, Office of Science, Office of Nuclear Physics under award number DE-FG02-87ER-40316. We also recognize support of co-author E. Aboud by the Nuclear Solutions Institute’s Graduate Student Merit Fellowship.

\section{Disclaimer}
\label{S:9}
This report was prepared as an account of work sponsored by an agency of the United States Government. Neither the United States Government nor any agency thereof, nor any of their employees, makes any warranty, express or implied, or assumes any legal liability or responsibility for the accuracy, completeness, or usefulness of any information, apparatus, product, or process disclosed, or represents that its use would not infringe privately owned rights. Reference herein to any specific commercial product, recommendation, or favoring by the United States Government or any agency thereof. The views and opinions of authors expressed herein do not necessarily state or reflect those of the United States Government or any agency thereof.

\bibliographystyle{ieeetr}

\bibliography{manuscript.bbl}

\end{document}